# Ferroelectric modulation of quantum emitters in monolayer WS$_2$

Sung-Joon Lee[a][†], Hsun-Jen Chuang[†], Andrew Yeats, Kathleen M. McCreary, Dante J. O'Hara, Berend T. Jonker*

U.S. Naval Research Laboratory, Washington, DC, 20375, USA

[a] Postdoctoral associate at the U.S. Naval Research Laboratory through the American Society for Engineering Education

[†] These authors contributed equally to this work.

* Email: berry.jonker@nrl.navy.mil

**Abstract**

Quantum photonics promises significant advances in secure communications, metrology, sensing and information processing/computation. Single photon sources are fundamental to this endeavor. However, the lack of high quality single photon sources remains a significant obstacle. We present here a new paradigm for the control of single photon emitters (SPEs) and single photon purity by integrating monolayer WS$_2$ with the organic ferroelectric polymer poly(vinylidene fluoride-*co*-trifluoroethylene) (P(VDF-TrFE)). We demonstrate that the ferroelectric domains in the P(VDF-TrFE) film control the purity of single photon emission from the adjacent WS$_2$. By switching the ferroelectric polarization, we reversibly tune the single photon purity between the semi-classical and quantum light regimes, with single photon purities as high as 94%. This provides another avenue for modulating and encoding quantum photonic information, complementing more complex approaches. This novel multidimensional heterostructure introduces a new avenue for control of quantum emitters by combining the nonvolatile ferroic properties of a ferroelectric with the radiative properties of the zero-dimensional atomic scale emitters embedded in the two-dimensional WS$_2$ semiconductor monolayer.

**Introduction**

Two dimensional (2D) materials provide solid state platforms for a wide spectrum of new phenomena with many avenues for technological applications,[1] including high performance transistors,[2] interactions in novel van der Waals heterostructures,[3–5] 2D magnetic ordering,[6] exciting spintronic[7] and multiferroic effects,[8] neuromorphic computing,[9] and quantum information.[10,11] These materials also serve as hosts for quantum emitters (QEs), often referred to as single photon emitters (SPEs),[12] which are essential building blocks for the photonic processes and circuits envisioned to enable quantum information processing and sensing .[10,11] Quantum photonic circuits are expected to offer performance and functionality not possible with classical light.[13,14] These applications impose many requirements on QE candidates, including deterministic creation and placement of the emitter, a high degree of single photon purity ranging from 90-100%, and a mechanism to control or modulate such emission.

The planar atomic layered character of 2D materials offers many advantages over the more well studied QE candidates such as InAs-based quantum dots or nitrogen vacancy centers in diamond.[10,11] They are relatively easy to synthesize, and lack the out-of-plane bonding typical of other materials, so that they can be readily integrated with a variety of substrates without the complications of lattice matching. In addition, the QE site is located very near the surface, thus avoiding losses from total internal reflection and providing high photon extraction efficiency. Surface proximity also make these QEs attractive for heterogeneous integration with different materials or architectures, for instance *via* evanescent coupling to waveguides in photonic integrated circuits (PICs).

Significant progress has been reported in deterministic creation and placement of QEs in 2D materials using local strain fields in monolayer transition metal dichalcogenides, including strain induced by holes in the substrate[15], nanopillars[16–19], and nanoindentation techniques[20,21]. Although relatively high single photon emission purity has been reported for select samples,[22] in many cases the purity is severely compromised by the emission of semi-classical light arising from a variety of material-dependent sources commonly associated with defect bound excitons which is spectrally degenerate with the quantum light and cannot be simply filtered out.[17,23–25]

In recent work, we successfully demonstrated a three-fold improvement in the purity of single photon emission for emitters in monolayer tungsten diselenide (WSe$_2$) by applying a gate voltage and corresponding electric field[26]. This resulted in field-induced dissociation of the defect bound excitons whose luminescence would otherwise contaminate the purity of the quantum emitters in the same spectral region. This approach also provided some degree of control over the activation and deactivation of select quantum emitters. However, upon removal of the gate bias, the achieved purity level reverts to the classical regime. Therefore, there is a clear need for a nonvolatile mechanism that can maintain high emission purity.

We report here the nonvolatile and reversible control of single photon emission purity in monolayer tungsten disulfide (WS$_2$) by integrating it with an organic ferroelectric polymer, poly(vinylidene fluoride-*co*-trifluoroethylene (P(VDF-TrFE)). We create an emitter in the WS$_2$, and are able to toggle the emission between high purity quantum light (characterized by sub-Poissonian statistics) and semi-classical light (characterized by Poissonian statistics) by switching the ferroelectric polarization of the P(VDF-TrFE) with a bias voltage. We further demonstrate that the monolayer WS$_2$ operates as a transparent top gate — application of a bias voltage to the WS$_2$ switches the ferroelectric polarization of the underlying P(VDF-TrFE) film. This approach enables nonvolatile modulation of the emission character by reversing the polarization of the ferroelectric domain under a given emitter, thereby switching between the quantum and semi-classical regimes. This provides another avenue for encoding quantum photonic information, complementing more complex approaches such as spectral shearing[27]. We achieve single photon emission purity ***P*** as high as 94%, as determined by the measured second order correlation function g$^{(2)}$(t=0) value of 0.06, where ***P*** = 1 - g$^{(2)}$(0), the highest purity reported for WS$_2$ SPEs to our knowledge[28,29]. This novel heterostructure introduces a new paradigm for control of quantum emitters by combining the nonvolatile ferroic properties of a ferroelectric with the radiative properties of the zero-dimensional atomic scale emitters embedded in the two-dimensional WS$_2$ semiconductor monolayer.

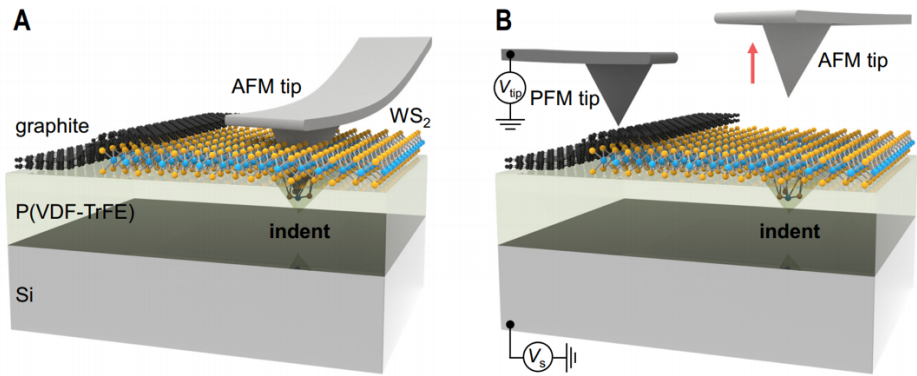

**Fig. 1. Illustration of the WS$_2$ single photon emitter on a ferroelectric P(VDF-TrFE)/Si substrate.** (**A**) Using the nanoindentation technique with an AFM tip, quantum emitters are created by the high local strain fields at specific idented positions on monolayer WS$_2$. (**B**) Applying bias with a conductive PFM tip situated on graphite, the ferroelectric P(VDF-TrFE) beneath the monolayer WS$_2$ becomes polarized, with the orientation of the polarization aligned along the surface normal and pointing either in or out of the surface, depending upon the tip bias.

The samples consist of monolayer films of WS$_2$ grown by chemical vapor deposition (CVD) and mechanically transferred onto a 260 nm film of P(VDF-TrFE), which had been previously transferred onto a highly doped (p$^{++}$) Si substrate. We deterministically create and place quantum emitters within the WS$_2$ using the atomic force microscope (AFM) nanoindentation technique described previously[20] and illustrated in Fig. 1A, where the P(VDF-TrFE) serves as a deformable polymer. When the AFM tip is removed, the WS$_2$ conforms to the contour of the nanoindent, and the local strain field activates single photon emission from atomic scale defect states in the WS$_2$. For the top electrical contact, graphite was then transferred and partially covered the WS$_2$, and a conductive piezo force microscopy (PFM) tip was used to apply a bias voltage to switch the polarization of the P(VDF-TrFE) beneath the WS$_2$, as shown in Fig. 1B. Achieving intimate contact between WS$_2$ and the ferroelectric P(VDF-TrFE) film is crucial, and requires an ultra-smooth ferroelectric film surface. Therefore, a spin-coating and flip-over process was used for the P(VDF-TrFE) film as described in the Supplementary Material and Figure S1.    With a coercive field of approximately 50 MV m$^{-1}$([30]), a 260 nm thick P(VDF-TrFE) film requires an applied voltage of order 10 V to reverse its spontaneous polarization. Details of the morphology and writing/imaging polarization domains directly in the P(VDF-TrFE) film are found in Supplementary Fig. S2 and S3.

To demonstrate switching of the ferroelectric film, we employed a conductive PFM tip in contact mode to apply a tip voltage ($V_{tip}$) of ±10 V to the graphite contact while simultaneously applying a sample voltage ($V_s$) of opposite polarity to the conductive Si substrate. This process produced polarization domains within the P(VDF-TrFE) film beneath the WS$_2$ flake. Subsequently, we utilized the PFM mode to visualize and characterize those domains. A negative tip voltage creates polarization dipoles along the surface normal pointing out of the sample ("up" domain), and a positive surface charge as illustrated in Fig. 2A. A positive tip voltage creates the opposite domain polarization and surface charge (Fig. 2B).

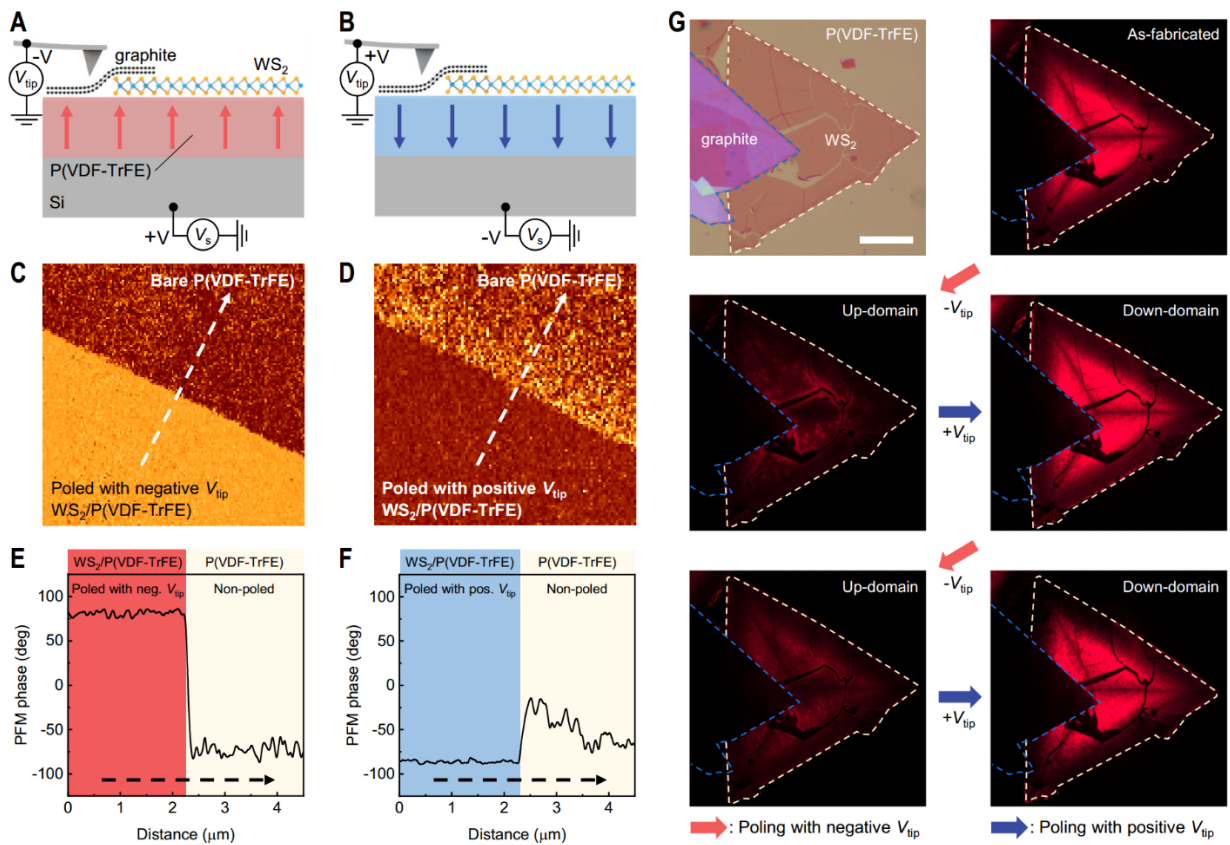

**Fig. 2. Characterization of WS$_2$ on P(VDF-TrFE) polarization domains.** (**A** and **B**) Cross-sectional schematics of WS$_2$/P(VDF-TrFE)/Si, illustrating the orientation of the polarization domains after negative (A) and positive (B) tip bias poling processes. (**C** and **D**) PFM images of poled WS$_2$/P(VDF-TrFE) surface and non-poled bare P(VDF-TrFE) surface, after negative (C) and positive (D) tip bias poling processes. The image size is 5 x 5 µm. (**E** and **F**) Horizontal PFM line scans along the dashed line, after negative (E) and positive (F) tip bias poling processes. (**G**) A bright field optical image of WS$_2$/P(VDF-TrFE) with graphite contact, and its sequential fluorescence images at room temperature, following each poling process. The location of graphite and monolayer WS$_2$ is indicated by blue and white dashed lines, respectively. Scale bar, 20 µm. All data obtained at 300K.

A PFM phase image of the WS$_2$/P(VDF-TrFE) poled with a negative $V_{tip}$ displays strong contrast between the poled and unpoled regions, signifying successful poling of the P(VDF-TrFE) film (Fig. 2C). A PFM line scan (depicted by the white dotted line in the phase image of Fig. 2C) across the edge of the WS$_2$ and onto the bare, unpoled P(VDF-TrFE) is shown in Fig. 2E. A phase change of over 150 degrees is measured, demonstrating that applying bias to the WS$_2$ flake successfully creates the ferroelectric domain underneath it, and not elsewhere.

A positive tip voltage is then applied to the same WS$_2$ to reverse the ferroelectric polarization, creating a polarization domain with dipoles facing into the sample ("down" domain) and a negative charge on the P(VDF-TrFE) surface (Fig. 2B). The corresponding phase image and line scan are shown in Fig. 2D and 2F, respectively. The large phase contrast observed between the "up" and "down" polarization domains shown in Fig. 2C and 2D demonstrates that the application of a bias to the WS$_2$ flake switches the ferroelectric domain between an upward and downward configuration.

Room temperature fluorescence (FL) images of the sample are shown in Fig. 2G for a sequence of negative-to-positive $V_{tip}$ poling cycles that repeatedly switch the ferroelectric polarization beneath the WS$_2$ between "up" and "down" domains. The luminescence of the WS$_2$ flake is strongly modulated by the orientation of the ferroelectric domain, as demonstrated previously[31]: the intensity is bright over a "down" domain in the underlying P(VDF-TrFE) and dark over an "up" domain. This luminescence can be altered in a reversible manner by applying an appropriate bias to the WS$_2$ flake. Notably, this procedure is in contrast with previous work where polarization domains were first written into a ferroelectric film (lead zirconium titanate, PZT) prior to transferring a WS$_2$ monolayer, and were not switched after this transfer[31].

The corresponding photoluminescence (PL) spectra of the monolayer WS$_2$ were measured at 300 K for the two different polarization domains, and provide evidence of charge doping. For the down-polarization domain (bright luminescence), the PL spectrum (solid black line) displays a prominent peak at around 2.01 eV, and high intensity (Fig. 3A). Conversely, the spectrum from the up-polarization domain shows a comparatively lower intensity, with a peak around 2.02 eV (Fig. 3B). These PL intensities mirror the modulation of the luminescence brightness shown in

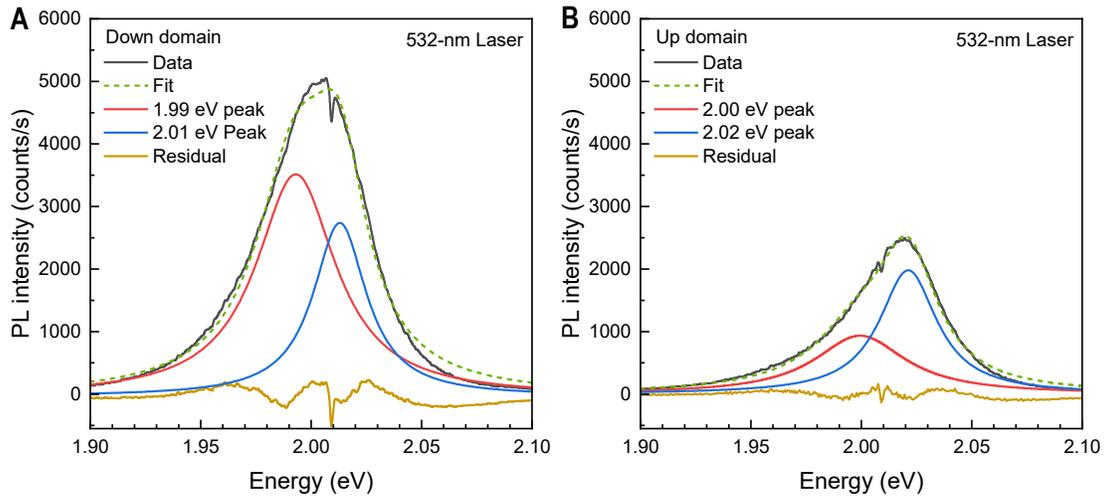

**Fig. 3. Photoluminescence (PL) spectra comparison of monolayer WS$_2$ on different P(VDF-TrFE) ferroelectric domains.** (**A**) Raw PL spectrum of monolayer WS$_2$ (black line) on P(VDF-TrFE) measured under the down-domain condition, fitted (dotted green line) with two Lorentzians centered at 1.99 eV (red line) and 2.01 eV (blue line). (**B**) Raw PL spectrum of monolayer WS$_2$ (black line) on P(VDF-TrFE) measured under the up-domain condition, fitted (dotted green line) with two Lorentzians centered at 2.00 eV (red line) and 2.02 eV (blue line). The residuals, representing the difference between the spectrum and the fit, are shown as yellow lines. All the data were taken at 300 K with 532nm continuous wave (CW) laser excitation and power density of 13.2 µW/µm$^2$.

Fig. 2G, further confirming the significant intensity differences observed in the WS$_2$ emission between the down- and up-polarization domains.

The PL spectra can be quantitatively fit employing two Lorentzian components (red and blue lines), resulting in excellent fits with low residual curves (green dashed and yellow lines, respectively) (Fig. 3A and 3B). The fitting process allowed the variation of parameters such as Lorentzian peak energy, height, and full width at half-maximum (FWHM), resulting in optimal peak energies of 1.99 and 2.01 eV for the down-polarization domain, and 2.00 and 2.02 eV for the up-polarization domain. These spectral features are attributed to the trion and free exciton, respectively, based on their energies and the energy difference[31]. The spectra from the down-polarization domain were dominated by the lower-energy trion component, identified as the positive trion due to the positive surface charge attracted by the ferroelectric polarization dipole. Conversely, the spectra from the up-polarization domain were dominated by the higher-energy free exciton component. These observations indicate that the polarization domain in the underlying P(VDF-TrFE) film strongly influences both the PL intensity and spectral distribution.

A proximal probe-based nano-indentation method[20] was employed to achieve deterministic creation and precise positioning of quantum emitters in monolayer $WS_2$ on the P(VDF-TrFE)/Si substrate. Controlled indents were made at specific locations using an AFM tip, (Fig. 1A), causing deformation in the $WS_2$/P(VDF-TrFE) heterostructure and generating a highly localized strain field which activates single photon emitters in the $WS_2$. The spatial fluorescence image of the sample (inset of Fig. 4) shows bright points of emission coinciding with most of the nanoindents in the heterostructure. As shown in Fig. 4, photoluminescence spectra obtained from such an indent exhibit sharp bright peaks characteristic of SPEs (red trace), while spectra obtained from non-indented areas of the WS$_2$ show only broad low intensity background in the same spectral region (black trace). All indented sites show sharp bright peaks consistent with SPEs. Because some indents exhibit more than one such emission site, making quantitative analysis of an individual emitter difficult within our spatial resolution, in the following we focus on those sites where one single emitter can be clearly identified and analyzed within the spatial resolution of our instrument.

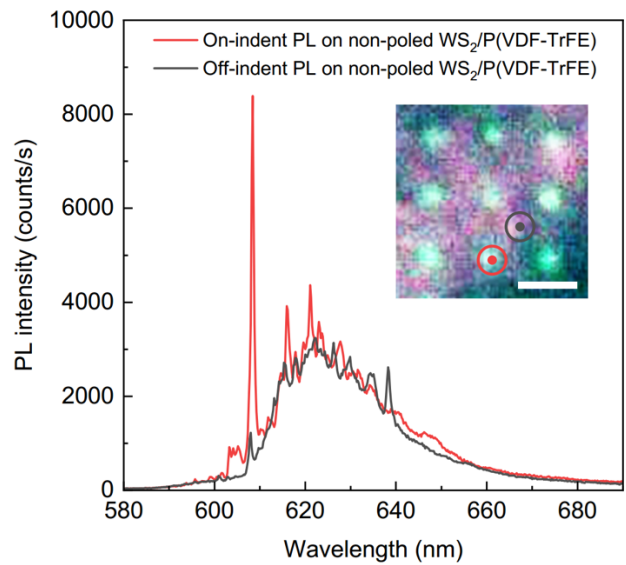

**Fig. 4. Localization of $WS_2$ single photon emitters using AFM indentation.** *Inset:* Fluorescence image showing that spatial positions of emission features correlate with positions of nanoindents. Scale bar, *5* μm. Photoluminescence spectra obtained from the monolayer $WS_2$ away from an indent (black curve, black circle on image) and on an indent (red curve, red circle on image) before poling the P(VDF-TrFE). The data were acquired at 5 K with 532nm CW laser excitation and power density of 0.4 μW/μm$^2$.

Subsequent application of either a negative or positive bias voltage via the PFM tip and graphite contact generates either an "up" or "down" polarization domain, respectively, in the P(VDF-TrFE) film under the WS$_2$ flake containing the indented region and emitter, as shown in Fig. 2 and discussed above. The poling process is accomplished by removing the sample from the cryostat, applying appropriate bias at room temperature in ambient with the PFM tip, returning the sample to the cryostat, and cooling to 5 K. This process sometimes resulted in small variations in emitter wavelength attributed to slight changes in the local strain field due to thermal cycling. Notably, the emitters were robust against such repeated thermal cycling and ambient exposure. Each emitter was identified and relocated by topographical features on the sample surface.

This ferroelectric domain orientation has a profound effect on the character of the light emitted from a given emitter, and we examine the corresponding photophysics by measuring the second order autocorrelation function, $g^{(2)}(t)$. A dip in $g^{(2)}(t)$ at zero time delay is referred to as antibunching, and indicates a reduced probability of detecting more than one photon. Values of $g^{(2)}(t=0) < 0.5$ indicate that the light originates from a quantum emitter, *i.e.* a single photon source rather than from a conventional source emitting "semi-classical" light such as typical electro- or photo-luminescence. For an ideal quantum emitter, $g^{(2)}(t=0) = 0$, *i.e.* the probability of simultaneously emitting two photons is zero.

PL spectra and $g^{(2)}(t)$ data obtained at 5 K are presented in Fig. 5 as the ferroelectric domain polarization is changed from "up" (Fig. 5, A and B) to "down" (Fig. 5, C and D), and back to "up" (Fig. 5, E and F). Each PL spectrum exhibits low background and is dominated by a narrow intense peak at ~607 nm of comparable intensity (within a factor of 3) with a full-width-at-half-maximum (FWHM) of ~0.6 nm, consistent with a single photon emitter. The Hanbury Brown and Twiss (HBT) methodology was used to acquire $g^{(2)}(t)$ data from the yellow highlighted region in each spectrum (see Methods for details). Both PL and $g^{(2)}(t)$ were acquired simultaneously in a series of exposures and then averaged using consistently-applied post-selection criteria described in the Methods. No background correction was applied to either the PL spectra or $g^{(2)}(t)$ data.

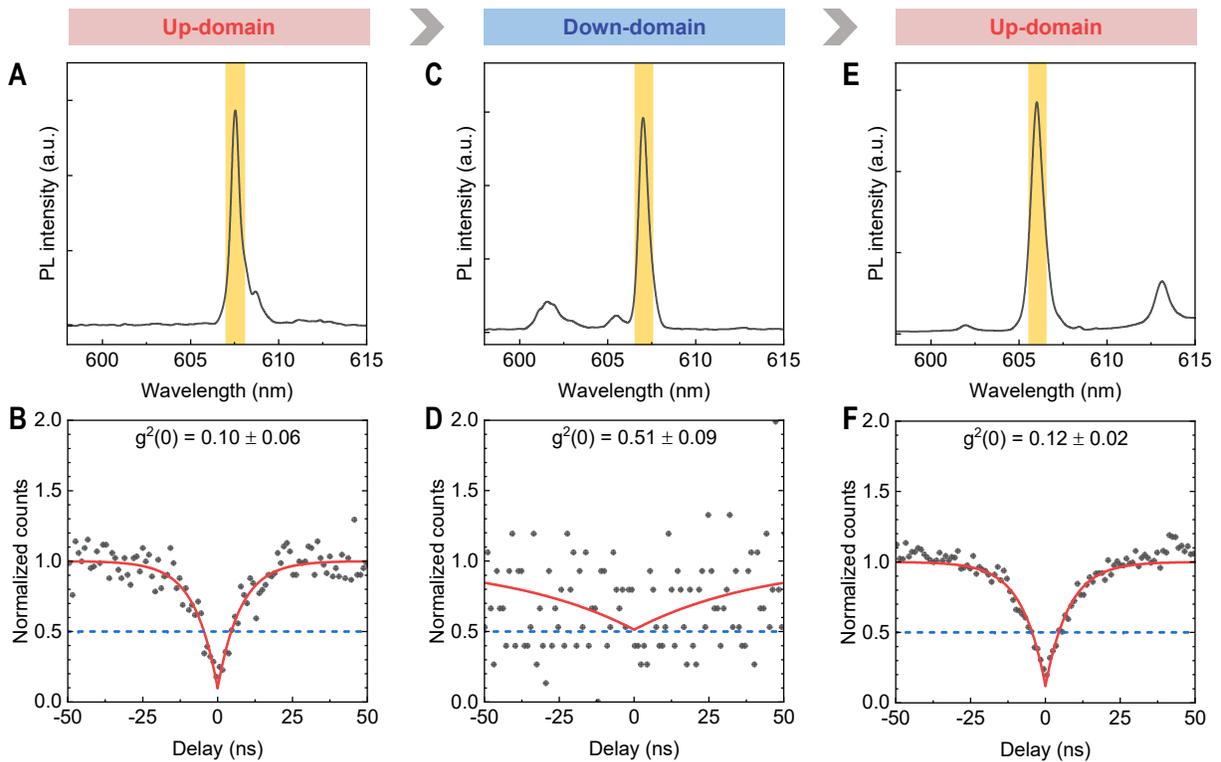

**Fig. 5. Optical characteristics of a ferroelectrically modulated WS$_2$ single photon emitter.**
(**A**) Photoluminescence (PL) spectrum of an indented WS$_2$ emitter on the up-domain P(VDF-TrFE). (**B**) Corresponding $g^{(2)}(t)$ plot with $g^{(2)}(t=0)$ value of 0.10 ± 0.06 , demonstrating that it is a high purity single photon emitter. The emitter lifetime obtained from these data is 7.7 ns. (**C**) PL spectrum of the same emitter after switching the orientation of the underlying P(VDF-TrFE) to the down-domain orientation. Note the appearance of a new feature at 602 nm. (**D**) Corresponding $g^{(2)}(t)$ plot with $g^{(2)}(t=0)$ value of 0.51 ± 0.09, indicating that the emission is now classical in character. The emitter lifetime obtained from these data is 43.4 ns. (**E**) PL spectrum of the same emitter after the polarization of the underlying P(VDF-TrFE) has been switched back to the up-domain orientation. (**F**) Corresponding $g^{(2)}(t)$ plot with $g^{(2)}(t=0)$ value of 0.12 ± 0.02 , demonstrating that the emitter returns to its original state as a high purity single photon emitter. The emitter lifetime obtained from these data is 7.9 ns, nearly identical to its original value. The yellow highlighted vertical bars in the PL spectra represent the effective 1 nm filter bandwidth used for $g^{(2)}(t)$ measurements. Black dots represent the data points, solid red lines show the fitting results, and dashed blue lines indicate $g^{(2)}(t) = 0.5$, the threshold value confirming a single photon emitter. All the data were acquired at 5 K.

For the initial "up" domain orientation (Fig. 5, A and B), we obtain a value $g^{(2)}(t=0) = 0.10 \pm 0.06$, the lowest reported to date for monolayer $WS_2$, confirming that in this configuration, the emitter is a source of high purity quantum light suitable for applications such as quantum cryptography/random number generation or quantum key distribution .[32] After reversing the P(VDF-TrFE) polarization to the "down" orientation (Fig. 5, C and D), we observe that an additional feature appears in the PL at a shorter wavelength (~602 nm) than the primary emitter peak (607 nm), indicating the formation of a state at higher energy from which radiative recombination is allowed. This is accompanied by a dramatic change in $g^{(2)}(t)$, with a much higher value $g^{(2)}(t=0) = 0.51 \pm 0.09$, indicating that the light originating from this emitter at 607 nm is no longer quantum in character, but in fact is now semi-classical light. After again reversing the P(VDF-TrFE) polarization to return it to the "up" orientation (Fig. 5, E and F), the PL peak at shorter wavelength is substantially reduced in intensity, and the measured $g^{(2)}(t=0) = 0.12 \pm 0.02$ for the prominent peak at 606 nm indicates that the emitter is again a source of high purity quantum light. The emitter lifetime extracted from the $g^{(2)}(t)$ data is 7.9 ns, similar to the original value of 7.7 ns measured in Fig. 5B, providing additional evidence that the emitter has not been altered and that the same emitting state is involved.

Thus, the orientation of the polarization domain in the P(VDF-TrFE) film determines the character of the light from the emitting state in the adjacent $WS_2$, enabling one to toggle back and forth from the quantum to semi-classical emission regimes. We observe this sequence of behavior for other emission sites where a single feature can be reliably measured, with single photon purities as high as 94% (Supplementary Fig. S4 and S5).

The standard model for a single photon emitter in a solid-state host is an atomic defect state within the host lattice with a simple two-level electronic structure, as shown in Fig. 6A, where radiative transitions are allowed between these two levels labeled "*E*" and "*G*". There is a non-zero time interval between the emission of subsequent photons from this state, because a cycle of excitation and relaxation must be completed between the two emission events, producing emission of discrete single photons. This results in the antibunching behavior which produces the dip in $g^{(2)}(t)$ at zero time delay. The measured emission behavior of the indented WS$_2$/P(VDF-TrFE) heterostructure for "up" ferroelectric domain orientation is consistent with this model, exhibiting high purity single photon emission.

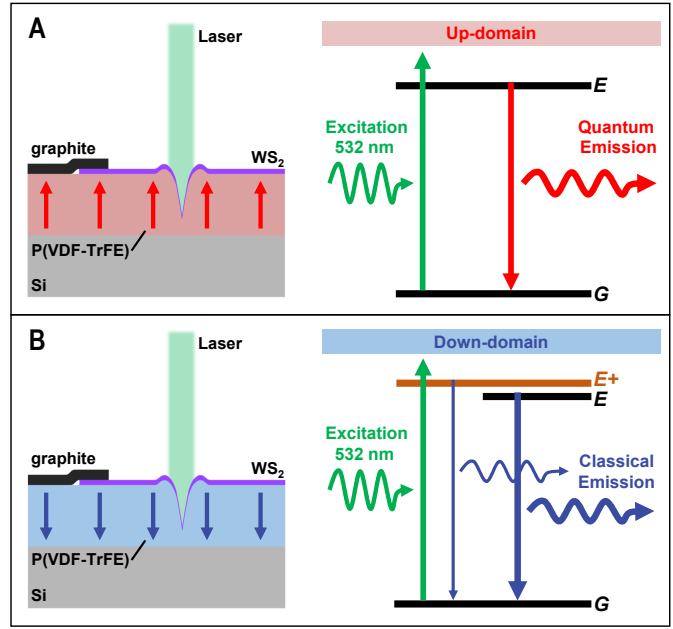

**Fig. 6. Schematic model of effect of ferroelectric domain orientation on WS$_2$ emission.** (**A**) For the up-domain orientation, the PL spectrum is dominated by a strong narrow peak at 607 nm, and high-purity single photon emission is observed, consistent with a simple two-level system. (**B**) For the down-domain orientation, an additional PL feature at shorter wavelength appears, signaling the presence of a new state, labeled *E+*, at higher energy, and the emitted light at 607 nm is now classical in nature.

When the ferroelectric polarization is switched to the "down" orientation, we observe the same dominant PL emission peak at ~607 nm (2.043 eV), but now consistently accompanied by a significant new feature at shorter wavelength (602 nm, 2.059 eV), revealing the formation of a new state at 16 meV higher energy. Surprisingly, the emitted light at 607 nm is now semi-classical in nature. We interpret this to mean that the "down" orientation transforms the electronic structure of the defect from a simple isolated two-level system to a more complex electronic structure which results in the emission of classical light. This is drawn for simplicity as a three-level system in Fig. 6B, with the new state at 16 meV higher energy labeled "*E+*", although the details are likely to be more complicated. This may be related to the charge doping expected for the "down" ferroelectric orientation, as indeed evidenced by the fact that trions dominate the PL spectrum, as shown in Fig. 2.

A first principles treatment of this problem is exceedingly challenging, because *(a)* the specific defect(s) giving rise to single photon emission in $WS_2$ is unknown, and *(b)* the optoelectronic properties of transition metal dichalcogenides (TMDs) such as $WS_2$ are well known to be dominated by excitonic behavior[33], and cannot be accurately captured by widely used computational tools such as standard density functional theory (DFT). This is in marked contrast to insulators such as diamond and hexagonal boron nitride, where single particle methodologies can readily calculate the properties of the various deep defect states ("color centers") which are candidates for single photon emission in these host materials. Modeling of TMD heterostructures is especially difficult[34]. In particular, the energy levels of defects in TMDs, which are the source of single photon emission, continue to be a topic of ongoing research, with some debate reflected in the current theoretical effort[35]. In lieu of a first principles model, we instead hypothesize that the *E+* level giving rise to the PL feature at 602 nm (2.059 eV) is a charged derivative of the original defect state producing the feature at 607 nm (2.043 eV). Charge fluctuations between *E* and *E+* on time scales much less than the 1-10 ns radiative lifetime compromise the character of the original two-level system producing single photon emission (Fig. 6A), resulting in a more complex electronic structure manifested by the emission of classical light.

In conclusion, we have demonstrated a new avenue for manipulation of quantum emitter defect states in solid state hosts such as monolayer $WS_2$ which provides control of their photophysical properties. Specifically, we show that light emission from $WS_2$ defect states can be toggled between quantum and classical regimes by reversal of the ferroelectric polarization in an adjacent layer, providing nonvolatile control of single photon emission. This modulation may offer an additional tool for secure communications and quantum encryption schemes based upon single photon sources. While a simplistic model consistent with the experimental data has been presented, a better physical understanding of the physical mechanisms would benefit from advances in first principles theoretical treatments which incorporate the excitonic character of the host TMD materials and interactions with the polarization in an adjacent ferroelectric layer. We hope that our results will stimulate theoretical efforts to elucidate the fundamental behaviors in such complex multidimensional hetero-bilayer structures and further advance the state-of-the-art for quantum science applications.

**Acknowledgements**

The research performed at the Naval Research Laboratory was supported by core programs. This research was performed while S.-J.L. held an American Society for Engineering Education fellowship at NRL.

**Contributions**

S.-J.L., H.-J.C., and B.T.J. conceived and designed the experiments. S.-J.L., H.-J.C., K.M.M., and D.J.O. fabricated the samples, and characterized them with PL, AFM and PFM. A.Y. performed PL and the time correlated single photon counting measurements. B.T.J., S.-J.L., and H.-J.C. wrote the manuscript. All authors discussed the results and commented on the manuscript.

**Ethics declarations**

The authors declare no competing interests.

**References**

(1) Lin, Y.-C.; Torsi, R.; Younas, R.; Hinkle, C. L.; Rigosi, A. F.; Hill, H. M.; Zhang, K.; Huang, S.; Shuck, C. E.; Chen, C.; Lin, Y.-H.; Maldonado-Lopez, D.; Mendoza-Cortes, J. L.; Ferrier, J.; Kar, S.; Nayir, N.; Rajabpour, S.; van Duin, A. C. T.; Liu, X.; Jariwala, D.; Jiang, J.; Shi, J.; Mortelmans, W.; Jaramillo, R.; Lopes, J. M. J.; Engel-Herbert, R.; Trofe, A.; Ignatova, T.; Lee, S. H.; Mao, Z.; Damian, L.; Wang, Y.; Steves, M. A.; Knappenberger, K. L. Jr.; Wang, Z.; Law, S.; Bepete, G.; Zhou, D.; Lin, J.-X.; Scheurer, M. S.; Li, J.; Wang, P.; Yu, G.; Wu, S.; Akinwande, D.; Redwing, J. M.; Terrones, M.; Robinson, J. A. Recent Advances in 2D Material Theory, Synthesis, Properties, and Applications. *ACS Nano* **2023**, *17* (11), 9694–9747. https://doi.org/10.1021/acsnano.2c12759.
(2) Kim, K.-H.; Oh, S.; Fiagbenu, M. M. A.; Zheng, J.; Musavigharavi, P.; Kumar, P.; Trainor, N.; Aljarb, A.; Wan, Y.; Kim, H. M.; Katti, K.; Song, S.; Kim, G.; Tang, Z.; Fu, J.-H.; Hakami, M.; Tung, V.; Redwing, J. M.; Stach, E. A.; Olsson, R. H.; Jariwala, D. Scalable CMOS Back-End-of-Line-Compatible AlScN/Two-Dimensional Channel Ferroelectric Field-Effect Transistors. *Nat.*

*Nanotechnol.* **2023**, *18* (9), 1044–1050. https://doi.org/10.1038/s41565-023-01399-y.

(3) Joshi, J.; Zhou, T.; Krylyuk, S.; Davydov, A. V.; Žutić, I.; Vora, P. M. Localized Excitons in $NbSe_2$-$MoSe_2$ Heterostructures. *ACS Nano* **2020**, *14* (7), 8528–8538. https://doi.org/10.1021/acsnano.0c02803.

(4) Joshi, J.; Scharf, B.; Mazin, I.; Krylyuk, S.; Campbell, D. J.; Paglione, J.; Davydov, A.; Žutić, I.; Vora, P. M. Charge Density Wave Activated Excitons in $TiSe_2$–$MoSe_2$ Heterostructures. *APL Mater.* **2022**, *10* (1), 011103. https://doi.org/10.1063/5.0067098.

(5) Taheri, M.; Brown, J.; Rehman, A.; Sesing, N.; Kargar, F.; Salguero, T. T.; Rumyantsev, S.; Balandin, A. A. Electrical Gating of the Charge-Density-Wave Phases in Two-Dimensional h-$BN$/$1T$-$TaS_2$ Devices. *ACS Nano* **2022**, *16* (11), 18968–18977. https://doi.org/10.1021/acsnano.2c07876.

(6) Gong, C.; Li, L.; Li, Z.; Ji, H.; Stern, A.; Xia, Y.; Cao, T.; Bao, W.; Wang, C.; Wang, Y.; Qiu, Z. Q.; Cava, R. J.; Louie, S. G.; Xia, J.; Zhang, X. Discovery of Intrinsic Ferromagnetism in Two-Dimensional van Der Waals Crystals. *Nature* **2017**, *546* (7657), 265–269. https://doi.org/10.1038/nature22060.

(7) Hoque, A. Md.; Zhao, B.; Khokhriakov, D.; Muduli, P.; Dash, S. P. Charge to Spin Conversion in van Der Waals Metal $NbSe_2$. *Appl. Phys. Lett.* **2022**, *121* (24), 242404. https://doi.org/10.1063/5.0121577.

(8) Gong, C.; Kim, E. M.; Wang, Y.; Lee, G.; Zhang, X. Multiferroicity in Atomic van Der Waals Heterostructures. *Nat. Commun.* **2019**, *10* (1), 2657. https://doi.org/10.1038/s41467-019-10693-0.

(9) Sadaf, M. U. K.; Sakib, N. U.; Pannone, A.; Ravichandran, H.; Das, S. A Bio-Inspired Visuotactile Neuron for Multisensory Integration. *Nat. Commun.* **2023**, *14* (1), 5729. https://doi.org/10.1038/s41467-023-40686-z.

(10) Liu, X.; Hersam, M. C. 2D Materials for Quantum Information Science. *Nat. Rev. Mater.* **2019**, *4* (10), 669–684. https://doi.org/10.1038/s41578-019-0136-x.

(11) Alfieri, A.; Anantharaman, S. B.; Zhang, H.; Jariwala, D. Nanomaterials for Quantum Information Science and Engineering. *Adv. Mater.* **2023**, *35* (27), 2109621. https://doi.org/10.1002/adma.202109621.

(12) Perebeinos, V. Two Dimensions and One Photon. *Nat. Nanotechnol.* **2015**, *10* (6), 485–486. https://doi.org/10.1038/nnano.2015.104.

(13) Nilsson, J.; Stevenson, R. M.; Chan, K. H. A.; Skiba-Szymanska, J.; Lucamarini, M.; Ward, M. B.; Bennett, A. J.; Salter, C. L.; Farrer, I.; Ritchie, D. A.; Shields, A. J. Quantum Teleportation Using a Light-Emitting Diode. *Nat. Photonics* **2013**, *7* (4), 311–315. https://doi.org/10.1038/nphoton.2013.10.

(14) Shomroni, I.; Rosenblum, S.; Lovsky, Y.; Bechler, O.; Guendelman, G.; Dayan, B. All-Optical Routing of Single Photons by a One-Atom Switch Controlled by a Single Photon. *Science* **2014**, *345* (6199), 903–906. https://doi.org/10.1126/science.1254699.

(15) Kumar, S.; Kaczmarczyk, A.; Gerardot, B. D. Strain-Induced Spatial and Spectral Isolation of Quantum Emitters in Mono- and Bilayer $WSe_2$. *Nano Lett.* **2015**, *15* (11), 7567–7573. https://doi.org/10.1021/acs.nanolett.5b03312.

(16) Li, H.; Contryman, A. W.; Qian, X.; Ardakani, S. M.; Gong, Y.; Wang, X.; Weisse, J. M.; Lee, C. H.; Zhao, J.; Ajayan, P. M.; Li, J.; Manoharan, H. C.; Zheng, X. Optoelectronic Crystal of Artificial Atoms in Strain-Textured Molybdenum Disulphide. *Nat. Commun.* **2015**, *6* (1), 7381. https://doi.org/10.1038/ncomms8381.

(17) Palacios-Berraquero, C.; Kara, D. M.; Montblanch, A. R.-P.; Barbone, M.; Latawiec, P.; Yoon, D.; Ott, A. K.; Loncar, M.; Ferrari, A. C.; Atatüre, M. Large-Scale Quantum-Emitter Arrays in Atomically Thin Semiconductors. *Nat. Commun.* **2017**, *8* (1), 15093. https://doi.org/10.1038/ncomms15093.

(18) Wu, W.; Dass, C. K.; Hendrickson, J. R.; Montaño, R. D.; Fischer, R. E.; Zhang, X.; Choudhury, T. H.; Redwing, J. M.; Wang, Y.; Pettes, M. T. Locally Defined Quantum Emission from Epitaxial Few-Layer Tungsten Diselenide. *Appl. Phys. Lett.* **2019**, *114* (21), 213102. https://doi.org/10.1063/1.5091779.

(19) Zhao, H.; Pettes, M. T.; Zheng, Y.; Htoon, H. Site-Controlled Telecom-Wavelength Single-Photon Emitters in Atomically-Thin $MoTe_2$. *Nat. Commun.* **2021**, *12* (1), 6753. https://doi.org/10.1038/s41467-021-27033-w.

(20) Rosenberger, M. R.; Dass, C. K.; Chuang, H.-J.; Sivaram, S. V.; McCreary, K. M.; Hendrickson, J. R.; Jonker, B. T. Quantum Calligraphy: Writing Single-Photon Emitters in a Two-Dimensional Materials Platform. *ACS Nano* **2019**, *13* (1), 904–912. https://doi.org/10.1021/acsnano.8b08730.

(21) So, J.-P.; Kim, H.-R.; Baek, H.; Jeong, K.-Y.; Lee, H.-C.; Huh, W.; Kim, Y. S.; Watanabe, K.; Taniguchi, T.; Kim, J.; Lee, C.-H.; Park, H.-G. Electrically Driven Strain-Induced Deterministic Single-Photon Emitters in a van Der Waals Heterostructure. *Sci. Adv.* **2021**, *7* (43), eabj3176. https://doi.org/10.1126/sciadv.abj3176.

(22) Kumar, S.; Brotóns-Gisbert, M.; Al-Khuzheyri, R.; Branny, A.; Ballesteros-Garcia, G.; Sánchez-Royo, J. F.; Gerardot, B. D. Resonant Laser Spectroscopy of Localized Excitons in Monolayer $WSe_2$. *Optica* **2016**, *3* (8), 882–886. https://doi.org/10.1364/OPTICA.3.000882.

(23) Srivastava, A.; Sidler, M.; Allain, A. V.; Lembke, D. S.; Kis, A.; Imamoğlu, A. Optically Active Quantum Dots in Monolayer $WSe_2$. *Nat. Nanotechnol.* **2015**, *10* (6), 491–496. https://doi.org/10.1038/nnano.2015.60.

(24) Koperski, M.; Nogajewski, K.; Arora, A.; Cherkez, V.; Mallet, P.; Veuillen, J.-Y.; Marcus, J.; Kossacki, P.; Potemski, M. Single Photon Emitters in Exfoliated $WSe_2$ Structures. *Nat. Nanotechnol.* **2015**, *10* (6), 503–506. https://doi.org/10.1038/nnano.2015.67.

(25) Dang, J.; Sun, S.; Xie, X.; Yu, Y.; Peng, K.; Qian, C.; Wu, S.; Song, F.; Yang, J.; Xiao, S.; Yang, L.; Wang, Y.; Rafiq, M. A.; Wang, C.; Xu, X. Identifying Defect-Related Quantum Emitters in Monolayer $WSe_2$. *Npj 2D Mater. Appl.* **2020**, *4* (1), 1–7. https://doi.org/10.1038/s41699-020-0136-0.

(26) Stevens, C. E.; Chuang, H.-J.; Rosenberger, M. R.; McCreary, K. M.; Dass, C. K.; Jonker, B. T.; Hendrickson, J. R. Enhancing the Purity of Deterministically Placed Quantum Emitters in Monolayer $WSe_2$. *ACS Nano* **2022**. https://doi.org/10.1021/acsnano.2c08553.

(27) Wright, L. J.; Karpiński, M.; Söller, C.; Smith, B. J. Spectral Shearing of Quantum Light Pulses by Electro-Optic Phase Modulation. *Phys. Rev. Lett.* **2017**, *118* (2), 023601. https://doi.org/10.1103/PhysRevLett.118.023601.

(28) Palacios-Berraquero, C.; Barbone, M.; Kara, D. M.; Chen, X.; Goykhman, I.; Yoon, D.; Ott, A. K.; Beitner, J.; Watanabe, K.; Taniguchi, T.; Ferrari, A. C.; Atatüre, M. Atomically Thin Quantum Light-Emitting Diodes. *Nat. Commun.* **2016**, *7* (1), 12978.

https://doi.org/10.1038/ncomms12978.

(29) Cianci, S.; Blundo, E.; Tuzi, F.; Pettinari, G.; Olkowska-Pucko, K.; Parmenopoulou, E.; Peeters, D. B. L.; Miriametro, A.; Taniguchi, T.; Watanabe, K.; Babinski, A.; Molas, M. R.; Felici, M.; Polimeni, A. Spatially Controlled Single Photon Emitters in hBN-Capped $WS_2$ Domes. *Adv. Opt. Mater.* **2023**, *11* (12), 2202953. https://doi.org/10.1002/adom.202202953.

(30) Hu, W. J.; Juo, D.-M.; You, L.; Wang, J.; Chen, Y.-C.; Chu, Y.-H.; Wu, T. Universal Ferroelectric Switching Dynamics of Vinylidene Fluoride-Trifluoroethylene Copolymer Films. *Sci. Rep.* **2014**, *4* (1), 4772. https://doi.org/10.1038/srep04772.

(31) Li, C. H.; McCreary, K. M.; Jonker, B. T. Spatial Control of Photoluminescence at Room Temperature by Ferroelectric Domains in Monolayer $WS_2$/PZT Hybrid Structures. *ACS Omega* **2016**, *1* (6), 1075–1080. https://doi.org/10.1021/acsomega.6b00302.

(32) Aharonovich, I.; Englund, D.; Toth, M. Solid-State Single-Photon Emitters. *Nat. Photonics* **2016**, *10* (10), 631–641. https://doi.org/10.1038/nphoton.2016.186.

(33) Qiu, D. Y.; da Jornada, F. H.; Louie, S. G. Screening and Many-Body Effects in Two-Dimensional Crystals: Monolayer $MoS_2$. *Phys. Rev. B* **2016**, *93* (23), 235435. https://doi.org/10.1103/PhysRevB.93.235435.

(34) Kundu, S.; Amit, T.; Krishnamurthy, H. R.; Jain, M.; Refaely-Abramson, S. Exciton Fine Structure in Twisted Transition Metal Dichalcogenide Heterostructures. *Npj Comput. Mater.* **2023**, *9* (1), 1–7. https://doi.org/10.1038/s41524-023-01145-x.

(35) Khalid, S.; Medasani, B.; Lyons, J. L.; Wickramaratne, D.; Janotti, A. The Deep-Acceptor Nature of the Chalcogen Vacancies in 2D Transition-Metal Dichalcogenides. *2D Mater.* **2024**, *11* (2), 021001. https://doi.org/10.1088/2053-1583/ad2108.

(36) Bune, A. V.; Fridkin, V. M.; Ducharme, S.; Blinov, L. M.; Palto, S. P.; Sorokin, A. V.; Yudin, S. G.; Zlatkin, A. Two-Dimensional Ferroelectric Films. *Nature* **1998**, *391* (6670), 874–877. https://doi.org/10.1038/36069.

(37) Tian, B. B.; Wang, J. L.; Fusil, S.; Liu, Y.; Zhao, X. L.; Sun, S.; Shen, H.; Lin, T.; Sun, J. L.; Duan, C. G.; Bibes, M.; Barthélémy, A.; Dkhil, B.; Garcia, V.; Meng, X. J.; Chu, J. H. Tunnel Electroresistance through Organic Ferroelectrics. *Nat. Commun.* **2016**, *7* (1), 11502. https://doi.org/10.1038/ncomms11502.

(38) Schellin, R.; Hess, G.; Kressman, R.; Wassmuth, P. Corona-Poled Piezoelectric Polymer Layers of P(VDF/TrFE) for Micromachined Silicon Microphones. *J. Micromechanics Microengineering* **1995**, *5* (2), 106. https://doi.org/10.1088/0960-1317/5/2/012.

(39) Bae, S.-H.; Kahya, O.; Sharma, B. K.; Kwon, J.; Cho, H. J.; Özyilmaz, B.; Ahn, J.-H. Graphene-P(VDF-TrFE) Multilayer Film for Flexible Applications. *ACS Nano* **2013**, *7* (4), 3130–3138. https://doi.org/10.1021/nn400848j.

(40) Genchi, G. G.; Ceseracciu, L.; Marino, A.; Labardi, M.; Marras, S.; Pignatelli, F.; Bruschini, L.; Mattoli, V.; Ciofani, G. P(VDF-TrFE)/$BaTiO_3$ Nanoparticle Composite Films Mediate Piezoelectric Stimulation and Promote Differentiation of SH-SY5Y Neuroblastoma Cells. *Adv. Healthc. Mater.* **2016**, *5* (14), 1808–1820. https://doi.org/10.1002/adhm.201600245.

(41) Wang, X.; Wang, P.; Wang, J.; Hu, W.; Zhou, X.; Guo, N.; Huang, H.; Sun, S.; Shen, H.; Lin, T.; Tang, M.; Liao, L.; Jiang, A.; Sun, J.; Meng, X.; Chen, X.; Lu, W.; Chu, J. Ultrasensitive and Broadband $MoS_2$ Photodetector Driven by Ferroelectrics. *Adv. Mater.* **2015**, *27* (42), 6575–

6581. https://doi.org/10.1002/adma.201503340.

(42)   Wu, G.; Tian, B.; Liu, L.; Lv, W.; Wu, S.; Wang, X.; Chen, Y.; Li, J.; Wang, Z.; Wu, S.; Shen, H.; Lin, T.; Zhou, P.; Liu, Q.; Duan, C.; Zhang, S.; Meng, X.; Wu, S.; Hu, W.; Wang, X.; Chu, J.; Wang, J. Programmable Transition Metal Dichalcogenide Homojunctions Controlled by Nonvolatile Ferroelectric Domains. *Nat. Electron.* **2020**, *3* (1), 43–50. https://doi.org/10.1038/s41928-019-0350-y.

(43)   Batool, S.; Idrees, M.; Han, S.-T.; Roy, V. A. L.; Zhou, Y. Electrical Contacts With 2D Materials: Current Developments and Future Prospects. *Small* **2023**, *19* (12), 2206550. https://doi.org/10.1002/smll.202206550.

(44)   Liu, X.; Lu, Y.; Yu, W.; Wu, J.; He, J.; Tang, D.; Liu, Z.; Somasuntharam, P.; Zhu, D.; Liu, W.; Cao, P.; Han, S.; Chen, S.; Seow Tan, L. AlGaN/GaN Metal-Oxide-Semiconductor High-Electron-Mobility Transistor with Polarized P(VDF-TrFE) Ferroelectric Polymer Gating. *Sci. Rep.* **2015**, *5* (1), 14092. https://doi.org/10.1038/srep14092.

(45)   Xia, W.; Peter, C.; Weng, J.; Zhang, J.; Kliem, H.; Jiang, Y.; Zhu, G. Epitaxy of Ferroelectric P(VDF-TrFE) Films via Removable PTFE Templates and Its Application in Semiconducting/Ferroelectric Blend Resistive Memory. *ACS Appl. Mater. Interfaces* **2017**, *9* (13), 12130–12137. https://doi.org/10.1021/acsami.7b01571.

(46)   Gu, J.; Wu, C.; He, X.; Chen, X.; Dong, L.; Weng, W.; Cheng, K.; Wang, D.; Chen, Z. Enhanced M2 Polarization of Oriented Macrophages on the P(VDF-TrFE) Film by Coupling with Electrical Stimulation. *ACS Biomater. Sci. Eng.* **2023**, *9* (5), 2615–2624. https://doi.org/10.1021/acsbiomaterials.2c01551.

(47)   Tsutsumi, N.; Hirano, Y.; Kinashi, K.; Sakai, W. Enhancement of Amplified Spontaneous Emission and Laser Performance of Rhodamine 6G/Cellulose Acetate DFB and DBR Waveguide Devices: A Role of Thermally Annealed P(VDF-TrFE) Intermediate Layer. *ACS Appl. Electron. Mater.* **2020**, *2* (6), 1514–1521. https://doi.org/10.1021/acsaelm.0c00124.

**Materials and Methods**

**Material growth of monolayer WS$_2$.**

Monolayer WS$_2$ is synthesized in a two-inch quartz tube furnace on SiO$_2$/Si substrates (275 nm oxide). To achieve a hydrophilic surface, the SiO$_2$/Si are soaked in a piranha etch composed of three parts concentrated sulfuric acid and one part hydrogen peroxide solution (30 wt.%) and subsequently rinsed in de-ionized water. Two growth substrates are loaded face down in a quartz boat that contains approximately 1 g of WO$_3$ powder (Alfa Aesar, 99.999% purity) and positioned in the center of the furnace for each growth run. Perylene-3,4,9,10-tetracarboxylic acid tetrapotassium salt molecules (1 mM concentration) are drop cast and dried on the upstream SiO$_2$/Si prior to loading into the furnace. A second boat containing sulfur powder (Alfa Aesar, 99.999% purity) is placed outside the central heating zone. Ultra-high purity argon (65 sccm) flows continuously through the furnace as it heats to the target temperature of 850 °C. Upon reaching the target temperature, 10 sccm of ultra-high purity H$_2$ is added to the Ar flow and maintained throughout the 10-minute soak and subsequent cooling to room temperature.

**Fabrication of the P(VDF-TrFE) thin film.**

A powder form of poly(vinylidene fluoride-*co*-trifluoroethylene) or P(VDF-TrFE) with a VDF/TrFE molar ratio of 70/30 was utilized for all experiments in this study. The P(VDF-TrFE) powder was dissolved in diethyl carbonate (DEC) to prepare a 2.5 wt.% solution, which involved mixing 0.4 g of P(VDF-TrFE) powder with 15.6 g of DEC. This solution was heated at 90 °C while stirred for 1 h to ensure complete dissolution. After cooling to room temperature, 50 µL of the prepared solution was spin-coated onto a 1 cm x 1 cm sacrificial SiO$_2$/Si substrate under atmospheric conditions, using a spin rate of 3000 rpm for 20 s. Based on our previous experience, each spin-coated layer achieved an approximate thickness of 65 nm. To reach the desired film thickness, a cyclic spin-coating process was employed. After each spin-coating cycle, the film was annealed at 90 °C for 1 min on a hot plate to facilitate solvent evaporation. Once the final layer was applied, the P(VDF-TrFE) films were carefully detached from the sacrificial SiO$_2$/Si substrate, flipped over, and physically laminated onto a designated conductive Si substate (Fig. S1).

Subsequently, the P(VDF-TrFE) films on the Si substrate were cured at 135 °C for 4 h to promote sufficient crystallinity.

**Fabrication of the WS$_2$/P(VDF-TrFE) with graphite contact.**

The monolayer WS$_2$ was carefully transferred onto a 260 nm thick P(VDF-TrFE) film positioned on a silicon (Si) substrate. In a concise summary of the procedure, the polydimethylsiloxane (PDMS) stamp was brought into contact with the WS$_2$ on the growth substrate (SiO$_2$/Si). A mixture of deionized water and 2-propanol in an approximate ratio of 2:1 was then applied to the sample. This caused the WS$_2$ monolayer to delaminate from the growth substrate and adhere to the PDMS stamp. Subsequently, the WS$_2$ on the PDMS stamp was dried using nitrogen gas. Finally, the monolayer WS$_2$ was brought into contact with the P(VDF-TrFE) surface, and the WS$_2$ was released from the PDMS onto the surface. For the fabrication of electrical contacts, graphite flakes were mechanically exfoliated onto a Piranha-cleaned SiO$_2$/Si wafer. This wafer was then subjected to the same transfer methods described above to stack the graphite flakes on top of the WS$_2$/P(VDF-TrFE) structure, facilitating the creation of electrical contacts.

**Fluorescence imaging.**

Fluorescence imaging was performed utilizing objectives with magnifications of either 50X or 100X. Our illumination source consisted of a LED fluorescence light system (X-Cite Xylis XT720S) equipped with a wideband green fluorescence filter cube (Olympus U-MWG2), which was integrated into the microscope setup (Olympus BX53M). For low-temperature measurement, fluorescence imaging was performed with mercury burner illumination source (USH-103OL), utilizing a long working distance objective with a 50X magnification.

**Indentation, poling process and characterization on AFM.**

AFM indentation of monolayer WS$_2$/P(VDF-TrFE) was conducted using a Park Systems NX10 AFM instrument. The indentations were performed utilizing the "Z scanner" mode within the nano-indentation mode of the Park NX10, allowing in-plane cantilever motion during the indentation process. In this study, we employed a PPP-NCHR cantilever for indentation. The poling process was carried out employing the PFM mode of the Park NX10. Polarization domains were induced within the P(VDF-TrFE) film using an PFM cantilever (Multi75-E, Cr/Pt-coated Si cantilever) by applying direct current (d.c.) voltages ranging from -10 V to +10 V to the AFM tip and substrate, $V_{tip}$ and $V_s$, respectively. To achieve up-domain polarization, the PFM cantilever was positioned on the graphite in contact on $WS_2$, after which a voltage of -10 V was applied to $V_{tip}$, while +10 V was applied to $V_s$, ensuring complete polarization of the P(VDF-TrFE) film. Conversely, for down-domain polarization, the voltage polarity was reversed. We employed the PFM mode with an oscillation drive amplitude of 2.5 V to visualize the polarization state of the poled regions at an oscillation modulation frequency of 34 kHz.

**Autocorrelation measurements.**

The second order autocorrelation function $g^{(2)}(t)$ was measured for each emitter using the standard Hanbury Brown and Twiss methodology using a 50:50 beam splitter, two Geiger mode Si avalanche photodiodes (Excelitas SPCM-AQRH-14), and time-correlated single photon counting electronics (PicoQuant Picoharp). The sample was cooled to 5 K in a closed-cycle optical cryostat with an *in situ* 100x 0.85 NA microscope objective. A scanning mirror and 4*f* relay lens system were used to map the surface reflectance and luminescence from the sample with < 1 µm resolution, allowing the same emitters to be located and re-measured after subsequent ferroelectric poling steps. Photoluminescence and $g^{(2)}(t)$ data were acquired using 532nm *cw* excitation with 20-50 nW power measured at the objective lens. Luminescence was separated from reflected pump light with a 600 nm long-pass dichroic mirror and then passed through a 600 nm long-pass filter (ThorLabs FELH0600) and a tunable short-pass filter adjusted to 613 nm (Semrock Versachrome TSP01-628). 1% of the collected and filtered light was diverted to a spectrometer for simultaneous monitoring of the emission spectrum (PI-Acton SP-2750i with PyLon 100BR_eX CCD). Both PL spectra and photon correlations were acquired simultaneously in a series of 120 s exposures and then averaged. To reduce noise from blinking and spectral wandering we excluded from our analysis exposures whose spectral maximum did not fall within the shaded region shown in each figure. Photon correlation histograms were generated using 256 ns time bins to maximize signal-to-noise without exceeding the expected timing jitter of the detectors. Values of $g^{(2)}(0)$ and lifetime $\tau$ were extracted from fits to the equation $g^{(2)}(t) = \left(\frac{n-1}{n}\right) + \frac{1}{n}\left(1 - e^{-(t-t_0)/\tau}\right)$.

# Supplementary Material: Ferroelectric modulation of quantum emitters in monolayer WS$_2$

Sung-Joon Lee[a][†], Hsun-Jen Chuang[†], Andrew Yeats, Kathleen M. McCreary, Dante J. O'Hara, Berend T. Jonker*

U.S. Naval Research Laboratory, Washington, DC, 20375, USA

[a]Postdoctoral associate at the U.S. Naval Research Laboratory through the American Society for Engineering Education

[†]These authors contributed equally to this work.

*Email: berry.jonker@nrl.navy.mil

**Fabrication and characterization of P(VDF-TrFE) films on Si**

P(VDF-TrFE) has gained increasing attention due to its outstanding ferroelectric and piezoelectric attributes, presenting advantages over conventional ferroelectric oxides that require high deposition temperatures. It typically exhibits a polarization in the range of 10 to 25 µC cm$^{-2}$ and possesses a coercive field ranging from 50 to 100 MV m$^{-1}$([36,37]). Its integration into devices such as electret microphones[38], flexible electronics[39], and biomedical equipment[40] highlights its significance in modern technology. Furthermore, these ferroelectric polymers can readily form heterostructures with 2D materials through robust van der Waals interfacial bonding[41,42], making it a pivotal component in the realm of materials science and quantum physics.

      To integrate the copolymer film onto a highly doped (p$^{++}$) silicon substrate, a combination of spin-coating and flip-over physical lamination techniques was employed (Fig. S1). Unlike P(VDF-TrFE) films produced solely through spin-coating, this approach yielded smooth film surfaces with a roughness (root mean square roughness, $R_q$) of about 2.2 nm (Fig. S2), resembling that of the sacrificial substrate. Achieving intimate contact between WS$_2$ and the ferroelectric P(VDF-TrFE) film is crucial for reliable electrical connections. Rough surfaces with uneven topography may introduce gaps or non-uniform contact, leading to intermittent or poor electrical connections[43]. Especially for the poling of the ferroelectric film through atomically flat WS$_2$, which forms an

interface with it, it is of utmost importance to ensure the smoothness of the P(VDF-TrFE) film. This smooth interface helps facilitate efficient polarization of the P(VDF-TrFE) film, enabling reliable and uniform polarization of the ferroelectric domain (Fig. S3). Therefore, the achieved smoothness of the P(VDF-TrFE) film plays a critical role in ensuring the successful poling of the ferroelectric domains and overall device performance.

In contrast to P(VDF-TrFE) films fabricated through conventional spin-coating[44–47], our methodology yielded film surfaces characterized by a low root mean square roughness ($R_q$) of approximately 2.2 nm, closely mirroring that of the sacrificial $SiO_2$/Si substrate (Fig. S2 and Table S1). To achieve this, the P(VDF-TrFE) films were initially fabricated on a sacrificial silicon dioxide using standard spin-coating procedures. Subsequently, they were mechanically peeled off, flipped over, and physically laminated onto the conductive silicon substate.

The presence of such a smooth interface greatly facilitates the efficient polarization of the P(VDF-TrFE) film, thereby enabling the consistent and uniform polarization of the ferroelectric domain, as exemplified in Fig. S3. Establishing intimate contact between $WS_2$ and ferroelectric P(VDF-TrFE) film is important for ensuring reliable effect of the ferroelectric domains. The presence of rough surfaces with non-uniform topography has the potential to introduce gaps or irregular contact points, thus predisposing to sporadic or suboptimal electrical connection, as substantiated by previous studies[43]. Consequently, the attained smoothness of the P(VDF-TrFE) film assumes a pivotal role in ensuring the successful poling of ferroelectric domain and, ultimately, significantly impacts the overall performance of the device.

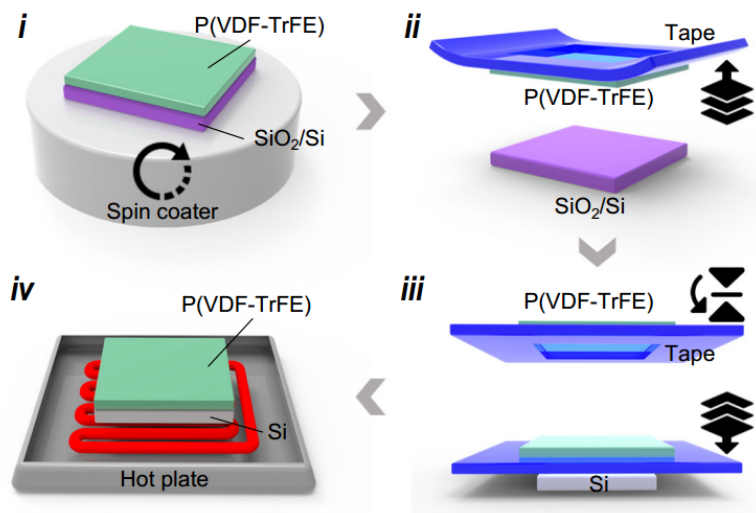

**Fig. S1. Fabrication process of P(VDF-TrFE)/Si substrate.** Schematic illustrations of fabrication steps for achieving a smooth film surface in the P(VDF-TrFE) film. The process involves spin-coating the P(VDF-TrFE) film onto a sacrificial SiO$_2$/Si substrate (**i**), followed by peeling-off (**ii**), flipping, and lamination onto a Si substrate (**iii**), and finally baking on a hot plate (**iv**).

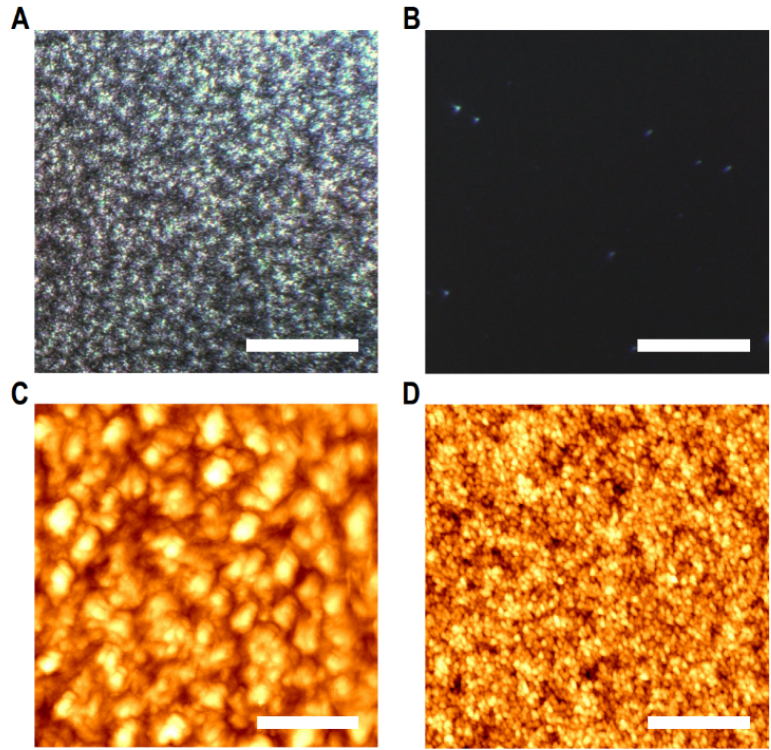

**Fig. S2. Surface morphology comparison and impact of flip-over physical lamination techniques on P(VDF-TrFE) film.** (**A** and **B**) Dark field images captured by optical microscopy of P(VDF-TrFE) film fabricated solely through spin-coating (A) and using physical lamination method (B). Scale bar, 40 μm. (**C**) AFM images of P(VDF-TrFE) film fabricated by spin-coating without flip-over physical lamination techniques, showing a roughness ($R_q$) of 15.8 nm. Scale bar, 2 μm. (**D**) AFM images of P(VDF-TrFE) film fabricated using physical lamination techniques showing a $R_q$ of 2.2 nm. Scale bar, 2 μm. Both cases have a consistent P(VDF-TrFE) film thickness of around 260 nm.

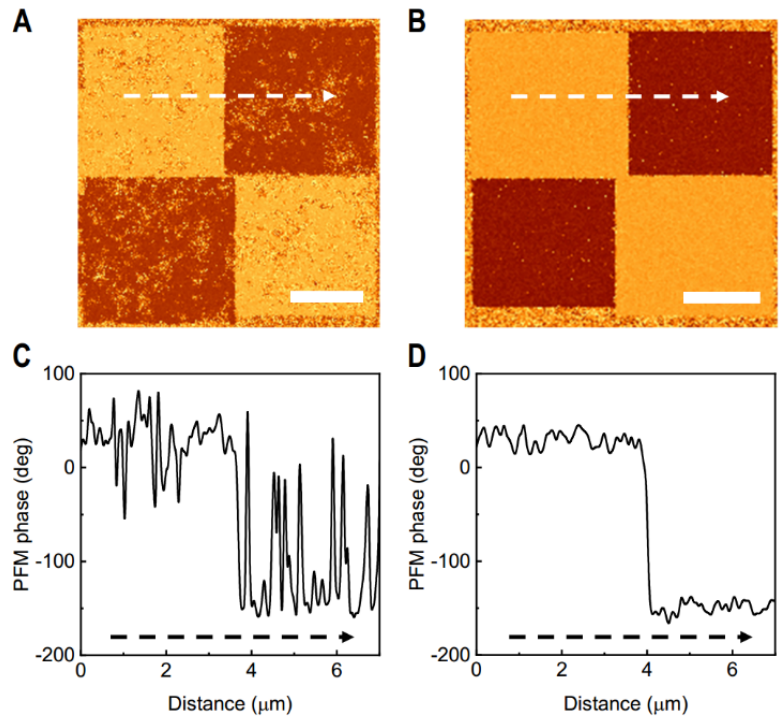

**Fig. S3. Impact of smoothness in P(VDF-TrFE) film for effective poling.** (**A**) PFM phase image of the poled P(VDF-TrFE) film with a roughness ($R_q$) of 23.11 nm. (**B**) PFM phase image of the poled P(VDF-TrFE) film with a $R_q$ of 2.37 nm. In the contact mode, a tip voltage of ±10 V was applied to the film surface, and 10 x 10 μm square polarization domains were written in a checkerboard pattern. The bright (dark) PFM boxes correspond to the areas induced by +10 V (-10 V) of a tip voltage. Scale bar, 5 μm. (**C**) Horizontal PFM line scan along the dashed line in (A), revealing the PFM phase of the P(VDF-TrFE) film with larger roughness. The phase distribution appears less continuous and uniform. (**D**) Horizontal PFM line scan along the dashed line in (B), showing the PFM phase of P(VDF-TrFE) film with smaller roughness. The phase distribution exhibits greater continuity and uniformity.

**Table S1. Surface roughness ($R_q$) comparison of P(VDF-TrFE) films fabricated via spin-coating.** Data from this work and references.

| Spin-coated P(VDF-TrFE) films | | |
|---|---|---|
| Thickness of film (nm) | Surface roughness (nm) | Reference |
| 260 | 2.2 | Present work |
| 500 | 5.0 | 44 |
| 270 | 8.1 | 45 |
| 550 | 20.8 | 46 |
| 180 | 39.6 | 47 |

**Fig. S4. Additional sample supporting Fig. 5.**

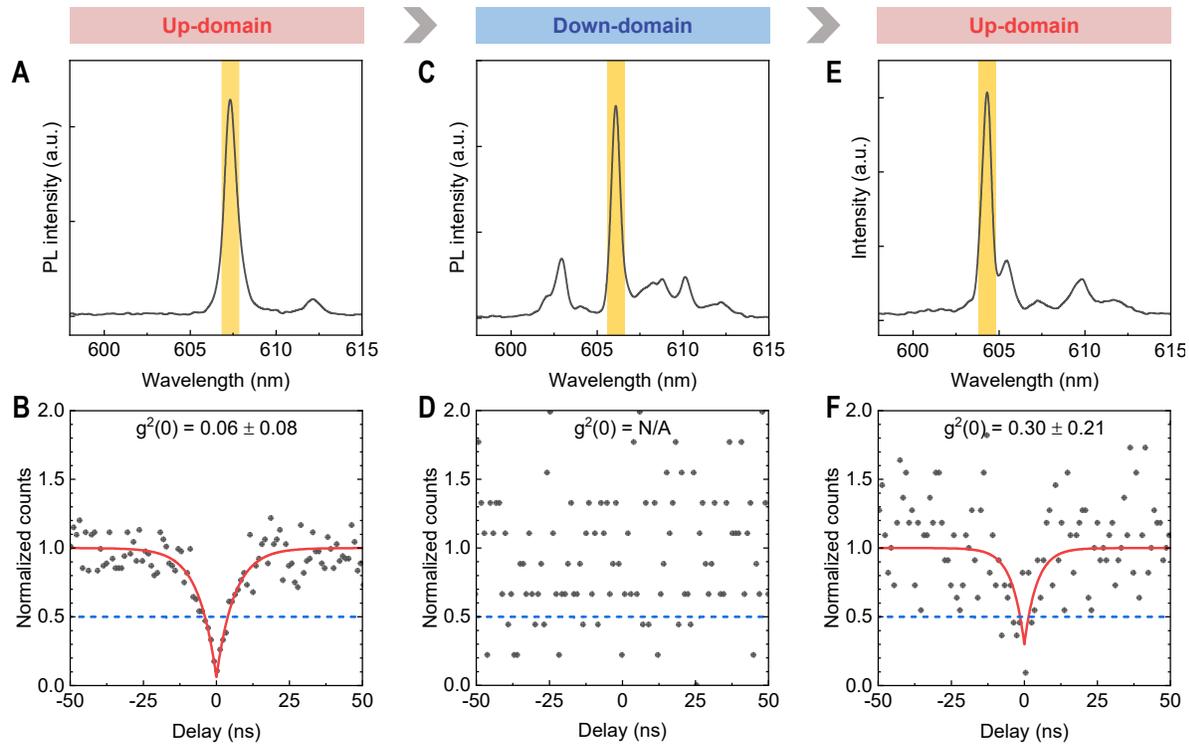

***Fig. S4. Additional WS$_2$ emitter exhibiting ferroelectrically modulated quantum emission character.***

(**A**) Photoluminescence (PL) spectrum of an indented WS$_2$ emitter on the up-domain P(VDF-TrFE). (**B**) Corresponding $g^{(2)}(t)$ plot with $g^{(2)}(t=0)$ value of 0.06 ± 0.08 , demonstrating that it is a high purity single photon emitter. The emitter lifetime obtained from these data is 6.2 ns. (**C**) PL spectrum of the same emitter after switching the orientation of the underlying P(VDF-TrFE) to the down-domain orientation. Note the appearance of a new feature at 602-603 nm. (**D**) Corresponding $g^{(2)}(t)$ data exhibit no antibunching, indicating that the emission is now classical in character. No fit is shown for clarity. (**E**) PL spectrum of the same emitter after the polarization of the underlying P(VDF-TrFE) has been switched back to the up-domain orientation. (**F**) Corresponding $g^{(2)}(t)$ plot with $g^{(2)}(t=0)$ value of 0.30 ± 0.21, demonstrating that the emitter returns to its original state as a high purity single photon emitter. The higher $g^{(2)}(t=0)$ value is attributed to spectral overlap with an adjacent emission feature which cannot be excluded by the 1 nm filter. The emitter lifetime obtained from these data is 4.3 ns, similar to its original value. The yellow highlighted vertical bars in the PL spectra represent the effective 1 nm filter bandwidth used for $g^{(2)}(t)$ measurements. Black dots represent the data points, solid red lines show the fitting results, and dashed blue lines indicate $g^{(2)}(t) = 0.5$, the threshold value confirming a single photon emitter. All the data were acquired at 5 K.

**Fig. S5. Additional sample supporting Fig. 5.**

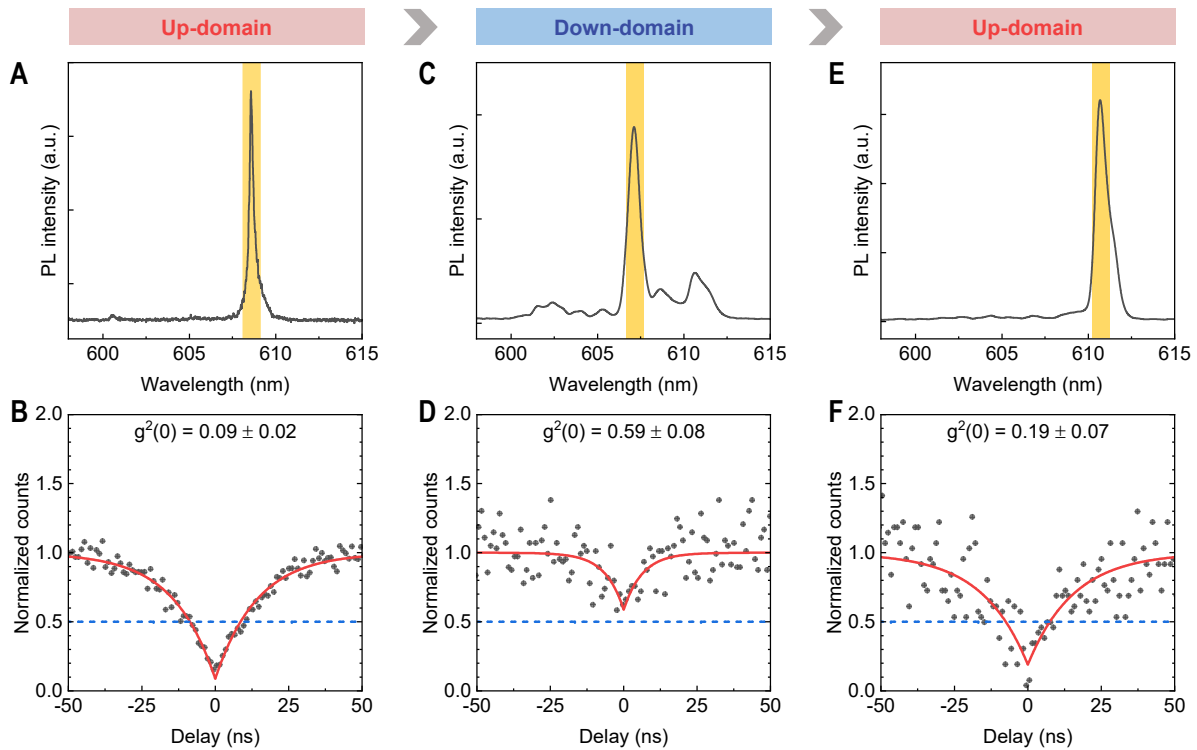

**Fig. S5. Additional WS$_2$ emitter exhibiting ferroelectrically modulated quantum emission character.**

(**A**) Photoluminescence (PL) spectrum of an indented WS$_2$ emitter on the up-domain P(VDF-TrFE). (**B**) Corresponding $g^{(2)}(t)$ plot with $g^{(2)}(t=0)$ value of 0.09 ± 0.02 , demonstrating that it is a high purity single photon emitter. The emitter lifetime obtained from these data is 14.3 ns. (**C**) PL spectrum of the same emitter after switching the orientation of the underlying P(VDF-TrFE) to the down-domain orientation. Note the appearance of a new feature at 602-603 nm. (**D**) Corresponding $g^{(2)}(t)$ plot with $g^{(2)}(t=0)$ value of 0.59 ± 0.08, indicating that the emission is now classical in character. The emitter lifetime obtained from these data is 6.3 ns. (**E**) PL spectrum of the emitter after the polarization of the underlying P(VDF-TrFE) has been switched back to the up-domain orientation. (**F**) Corresponding $g^{(2)}(t)$ plot with $g^{(2)}(t=0)$ value of 0.19 ± 0.07, demonstrating that the emitter returns to its original state as a high purity single photon emitter. The higher $g^{(2)}(t=0)$ value is attributed to spectral overlap with an adjacent emission feature which cannot be excluded by the 1 nm filter. The emitter lifetime obtained from these data is 15.5 ns, nearly identical to its original value. The yellow highlighted vertical bars in the PL spectra represent the effective 1 nm filter bandwidth used for $g^{(2)}(t)$ measurements. Black dots represent the data points, solid red lines show the fitting results, and dashed blue lines indicate $g^{(2)}(t)$ = 0.5, the threshold value confirming a single photon emitter. All the data were acquired at 5 K.